# Study on multi-fold bunch splitting in a high-intensity medium-energy proton synchrotron

Linhao Zhang[1,2,3], Min Chen[2], and Jingyu Tang[1,2,3,*]

[1]*Institute of High Energy Physics, CAS, Yuquan Road 19B, Beijing 100049, China*
[2]*University of Chinese Academy of Sciences, CAS, Yuquan Road 19A, Beijing 100049, China*
[3]*Spallation Neutron Source Science Center, Zhongziyuan Road NO. 1, Dongguan 523803, China*

Bunch splitting is an RF manipulation method of changing the bunch structure, bunch numbers and bunch intensity in the high-intensity synchrotrons that serve as the injector for a particle collider. An efficient way to realize bunch splitting is to use the combination of different harmonic RF systems, such as the two-fold bunch splitting of a bunch with a combination of fundamental harmonic and doubled harmonic RF systems. The two-fold bunch splitting and three-fold bunch splitting methods have been experimentally verified and successfully applied to the LHC/PS. In this paper, a generalized multi-fold bunch splitting method is given. The five-fold bunch splitting method using specially designed multi-harmonic RF systems was studied and tentatively applied to the medium-stage synchrotron (MSS), the third accelerator of the injector chain of the Super Proton-Proton Collider (SPPC), to mitigate the pileup effects and collective instabilities of a single bunch in the SPPC. The results show that the five-fold bunch splitting is feasible and both the bunch population distribution and longitudinal emittance growth after the splitting are acceptable, e.g., a few percent in the population deviation and less than 10% in the total emittance growth.

## I. INTRODUCTION

Beam manipulation in both the transverse and longitudinal phase space is a powerful technique to attain desirable beam properties in synchrotrons, such as bunch length and energy spread, emittance, bunch distribution and so on [1]. A variety of beam manipulation methods in the longitudinal phase space, including phase-space stacking, debunching, bunch-to-bucket transfer, bunch rotation, bunch merging or splitting, etc., have been applied by employing special settings of RF systems in many accelerators [2, 3]. In particular, bunch splitting is an RF manipulation method to change the bunch characteristics, like bunch numbers, bunch intensity, and the time structure of beam, in high-intensity synchrotrons serving as the injector of a particle collider.

In intense proton synchrotrons, two methods of splitting bunches have been utilized to obtain the desirable bunch properties [3, 4]. One conventional method for changing the bunch structure is to debunch the beam by turning off the RF voltage at the initial frequency, then to let the beam circulate without longitudinal RF focusing, and finally to rebunch the beam with another RF frequency. The other one is to adopt the combination of different harmonic RF systems, e.g., the two-fold splitting of a bunch with a combination of fundamental harmonic and doubled harmonic RF systems, which can minimize the longitudinal emittance growth and beam loss as compared with the debunching method. The bunch splitting method has been successfully applied to the LHC/PS for double splitting and triple splitting of one bunch [5, 6]. It has also been proposed and studied through the simulation in the JLEIC (Jefferson Lab Electron-Ion Collider) [7].

---

[*] Corresponding author.
tangjy@ihep.ac.cn

SPPC, as the second stage of the CEPC-SPPC (Circular Electron Positron Collider & Super Proton-Proton Collider) project, aims at exploring new physics beyond the Standard Model. The design luminosity at SPPC reaches $1\times10^{35}$ $cm^{-2}s^{-1}$ with the nominal bunch spacing of 25 ns, which will produce a large number of events per bunch crossing and thus impose a severe challenge on the detector trigger system and data analysis methods. To mitigate the pileup effects and the collective instabilities of a single bunch at SPPC, the operation mode of shorter bunch spacing, like 5 ns, was proposed at the CEPC Conceptual Design Report [8]. Likewise, the possibility of shorter bunch spacing has also been considered in the FCC-hh (Future Circular Collider for hadron-hadron collisions) project [9]. Therefore, it is inevitable and crucial to investigate the five-fold bunch splitting method to alter the bunch separation from 25 ns to 5 ns. This manipulation will be conducted in the Medium-Stage Synchrotron (MSS), the third accelerator of the SPPC injector chain.

The paper is organized as follows. The approaches of bunch splitting and the general multi-harmonic RF bunch splitting method are described in Sec. II. In Sec. III, the simulation method of bunch splitting and the application of the five-fold bunch splitting method to the MSS are presented. Conclusions and discussion are given in Sec. IV.

## II. BUNCH SPLITTING METHODS

There are two different bunch splitting methods used in synchrotrons: one is the debunching and rebunching method; the other is the multi-harmonic RF bunch splitting method. Both methods will have to be implemented under the stationary condition where the beam energy and magnetic field are kept constant to achieve the uniformity and symmetry of the bunch population after the splitting and relieve longitudinal emittance growth even the beam loss. The details of the two bunch splitting techniques are described below.

### A. Debunching and rebunching method

Debunching and rebunching is the most routine bunch manipulation method to change the number of bunches and the bunch spacing [10]. Rebunching is the reverse process of debunching. The procedures of this method are as follows: First, the RF voltage with the original harmonic frequency is adiabatically lowered to a level where the momentum spread is relatively low, then the RF system is turned off. Next, due to the turning-off of longitudinal focusing, particles with different momenta drift with different revolution frequencies, then the beam will fill up the whole ring and become the so-called coasting beam. Finally, a new RF system with the desirable harmonic frequency is turned on, and then the RF voltage is adiabatically increased to an appropriate value, during which the beam progressively receives an azimuthal modulation of density and is finally fully bunched in the new harmonic RF system.

To minimize the longitudinal emittance growth, the processes of debunching and rebunching have to be adiabatic or quasi-adiabatic. Even so, the increase of emittance is unavoidable since: (1) the RF voltage cannot be lowered down to a tiny amplitude, as the strong beam loading effect may occur; (2) during the drifting when the beam is left uncontrolled in the longitudinal plane, the space charge may introduce nonlinear effects and instabilities; (3) the nonlinear focusing field of the new RF system that is added suddenly will render the filamentation of the bunch in the phase space; (4) the coasting beam has a very small momentum spread which leads to the possible longitudinal microwave instability. In addition to the above drawbacks, as all the buckets are filled with bunches, the rebunched beam may not leave a sufficient time gap for the extraction kickers, which will result in intolerable beam loss during the extraction in high-intensity synchrotrons. Therefore, an alternative bunch splitting approach was proposed [4], which will be summarized in Sec. II.B.

## B. Multi-harmonic RF bunch splitting method

The main differences of the multi-harmonic RF bunch splitting method from the debunching-rebunching method lie in: (1) during the splitting, multiple harmonic frequencies coexist in the RF systems rather than appear successively; (2) there is not a phase of the full missing of RF focusing; (3) the gaps between different bunch trains can be preserved. These make the multi-harmonic RF bunch splitting method more suitable for bunch splitting in the injector chain for a hadron collider. In the following, as the most straightforward example, the two-fold bunch splitting that means one bunch splits into two bunches is introduced to illustrate how this method works, and then more general multi-harmonic RF bunch splitting will be discussed.

### 1. Two-fold bunch splitting with dual RF systems

Figure 1 illustrates the process of two-fold bunch splitting. First, one bunch is bound in the stationary single-harmonic RF bucket with a harmonic number $h_0$, see Fig. 1(a). Next, the second-harmonic RF system with a doubled frequency or harmonic number $2h_0$ is turned on to stretch the bunch, see Fig. 2(b). With the increase of the second-harmonic RF voltage, the bucket center gradually becomes the unstable fixed point from the stable one, and two new stable fixed points emerge on both sides of the bucket center. Then the bunch gets invaginated in the bucket center, and two new bunches are formed, encircling the two new stable fixed points, see Fig. 1(c)(d).

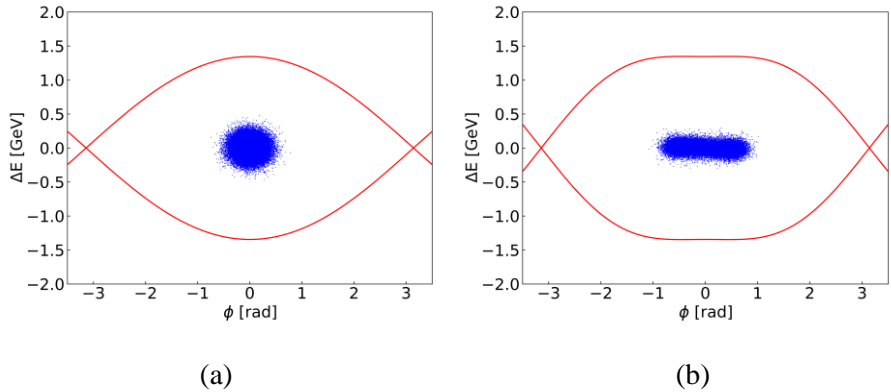

(a)                                          (b)

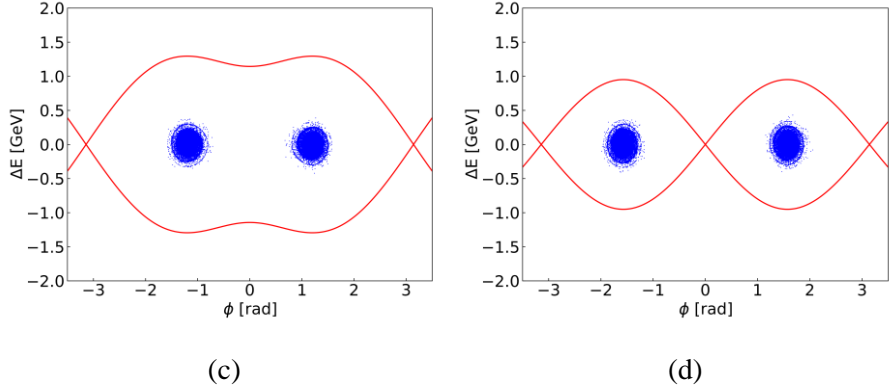

(c)                          (d)

FIG. 1. The schematics of the two-fold bunch splitting process.

To perform ideal bunch splitting, the phase difference between the two RF systems should be $\pi$, and the RF voltages should be programmed to follow the required pattern, which is shown in Fig. 2. One can see that in the first half of the process, the gradual increase of the doubled-harmonic RF voltage is to elongate the bunch, and in the second half, the gradual decrease of the fundamental harmonic RF voltage helps divide the bunch into two smaller ones.

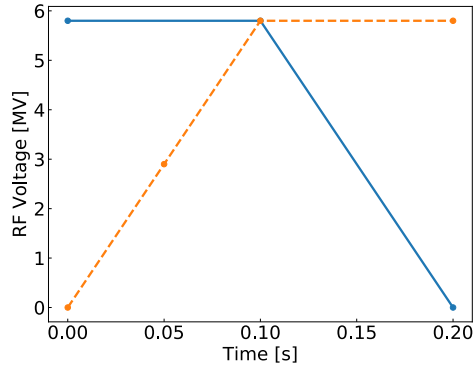

FIG. 2. The RF program during the process of two-fold bunch splitting. Blue solid line: fundamental harmonic RF voltage; orange dashed line: doubled-harmonic RF voltage.

### *2. Multi-bunch splitting with multiple harmonic RF systems*

Based on the two-fold bunch splitting experience described above and the three-fold bunch splitting experience in the LHC/PS, here we try to give the general method of one bunch splitting into multiple bunches using multiple harmonic RF systems. The bunch splitting method can be described as follows.

(1) The required number of harmonic RF systems depends on how many bunches one bunch is to be split into. For instance, for one bunch into three, three simultaneous RF systems with harmonic numbers $h_0$, $2h_0$, and $3h_0$ are needed. It is worth noting that the presence of the RF system with $2h_0$ is crucial, which assists in elongating the bunch before splitting and producing three more-or-less uniformly populated sub-bunches.

(2) To generate equal bunch splitting, the relative RF phases among the required different harmonic systems are definite. For bunch splitting in a synchrotron working below the transition energy, the phases for RF systems with even multiples of $h_0$ must be $\pi$, whereas those for RF systems with odd multiples of $h_0$ must be zero. The opposite is true for the case above transition energy.

(3) The voltages for different harmonic RF systems play critical roles during the bunch splitting process. It is impossible to derive analytical solutions to the RF voltage programs of different harmonic RF systems. However, one can choose the voltages at some pivotal moments and finally determine them based on the numerical simulations. Therefore, the RF voltage programs are not unique and are usually determined by trials. However, the time-dependent trends of the voltages for different harmonic RF systems are regular. The RF voltage with harmonic number $h_0$ first keeps constant and then needs to decrease to zero gradually. The voltages for the other harmonics should be increased progressively from zero in sequential order of the harmonics and reach their respective peaks successively. Then the voltages for the intermediate harmonics are decreased to zero in succession. Only the voltage with the highest harmonic remains in the last period of the splitting.

It is worth pointing out that the above rule is more suitable for the cases where the number of sub-bunches is a prime number. If the number of sub-bunches is a composite number, one can manipulate the fundamental bunch splitting more than once. For example, if one wants to split one bunch into four bunches, one can simply implement two-fold bunch splitting twice.

## 3. Key points for the RF voltage design

As partially mentioned above, to generate a good multi-fold bunch splitting with uniform particle partition and small emittance growth, the most essential and challenging task is to design the voltage patterns for the multiple harmonic RF systems. Besides the time-dependent trends for the multiple harmonic RF voltages mentioned above, another two key points also contribute to obtaining a better bunch splitting.

The first one is that the designed RF voltage scheme meets the adiabatic or quasi-adiabatic condition, which means the change of RF voltages must be sufficiently slow so that the redistribution of particles within the bucket is executed with an emittance growth as small as possible. The condition for adiabatic synchrotron motion can be evaluated with the adiabaticity coefficient $\alpha_{ad}$ as [2]:

$$\alpha_{ad} = \left| \frac{1}{\omega_s^2} \frac{d\omega_s}{dt} \right| = \frac{1}{2\pi} \left| \frac{dT_s}{dt} \right| \ll 1 . \qquad (1)$$

Here, $\omega_s$ and $T_s$ are the angular frequency and period of the synchrotron oscillation, respectively. Typically, the process is considered adiabatic when $\alpha_{ad} < 0.05$. This adiabatic condition is necessary to avoid the significant emittance growth by the filamentation effect of the particle distribution in the presence of a strong nonlinear RF field.

Another key point is to make the bunch stretched as far as possible before splitting the bunch or the intermediate bunch. Here, a general condition to flatten the bunch can be given based on the experience from the one in a dual-harmonic RF system [11]. A generalized RF voltage $V(\phi)$ seen by the beam at a certain moment, which is the summation of different harmonic RF voltages, can be expressed as:

$$V(\phi) = V_0 \sum_{i=1}^{n} k_i \sin(i\phi + \phi_i) \qquad (2)$$

where $n$ is the number of RF harmonics, $\phi$ is the phase in the fundamental RF, $\phi_i$ is the phase difference of the $i$-th harmonic RF to the fundamental RF and thus $\phi_1=0$, $V_0$ is the voltage of the fundamental RF at the initial moment, $k_i$ is the voltage ratio of the $i$-th harmonic RF to the fundamental RF at the initial moment. The corresponding potential well function can be expressed by:

$$U(\phi) = \frac{1}{V_0}\int_{\phi_s}^{\phi}[V(\phi)-V(\phi_s)]d\phi$$
$$= \sum_{i=1}^{n}\frac{k_i}{i}[\cos(i\phi_s+\phi_i)-\cos(i\phi+\phi_i)]-(\phi-\phi_s)\sum_{i=1}^{n}k_i\sin(i\phi_s+\phi_i) \quad , \quad (3)$$

where $\phi_s$ is the synchronous phase. To maximize the bunch length and obtain a flattened potential well, the derivatives of $V(\phi)$ from the first order to the $n$-th order should vanish at the center of the bunch:

$$\left.\frac{\partial V(\phi)}{\partial \phi}\right|_{\phi=\phi_s} = 0, \quad (4)$$

$$\ldots\ldots$$

$$\left.\frac{\partial^n V(\phi)}{\partial \phi^n}\right|_{\phi=\phi_s} = 0. \quad (5)$$

Then, one can obtain $\phi_{odd}=0$ when $i$ is odd, while $\phi_{even}=0$ when $i$ is even, and different $k_i$ values for all the harmonics. The highest order of the derivative is determined by the number of the required equations to solve all the $k_i$ values. Figure 3 shows the flattened potential wells obtained by using Eqs. (3)-(5) for $n=2$ ($k_2=0.5$), $n=3$ ($k_2=0.8$, $k_3=0.2$), and $n=5$ ($k_2=8/7$, $k_3=9/14$, $k_4=4/21$ and $k_5=1/42$). One can see that with more RF harmonics, one can obtain relatively more flattened or wider potential well and thus longer bunch. It should be pointed out that the above conditions are necessary to ensure no emergence of small inner buckets within the original outer bucket. In practice, to generate the desired particle partition in the higher-fold bunch splitting, one can adjust different $k$ values to make the initial bunch even more elongated than in the case of lower-fold bunch splitting.

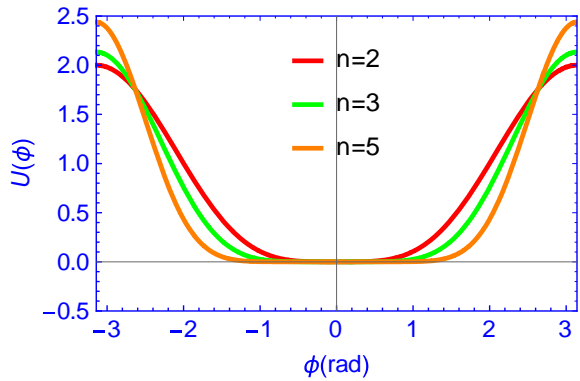

FIG. 3. The flattened potential well for multi-harmonic RF systems with the harmonics $n=2$ (red), $n=3$ (green) and $n=5$ (orange).

### III. APPLICATION OF THE FIVE-FOLD BUNCH SPLITTING METHOD TO THE MSS

The nominal bunch spacing in the SPPC baseline design is 25 ns. There are 10080 bunches in total for one of the two collider rings, each containing $1.5\times10^{11}$ particles, to achieve the nominal luminosity of $1\times10^{35}$ $cm^{-2}s^{-1}$. The instant luminosity could be even higher when some luminosity leveling methods are applied [12]. This will pose severe challenges on the detector design and data analysis methods that will allow us to record the data first and extract the interesting physics events afterwards. A natural idea to decrease the number of events per bunch collision is to reduce the bunch intensity by splitting one bunch into several bunches, which also means the reduction of bunch spacing. Although the instant luminosity is reduced with a smaller bunch population or bunch spacing and with the same total protons, the bunch spacing of 5 ns is still considered a good compromise. At SPPC, the solution to change the bunch spacing from the nominal 25 ns to the optional 5 ns is to perform the five-fold bunch splitting in the MSS, the third-stage accelerator of the SPPC injector chain. Likewise, the possibility of 5 ns bunch spacing has also been considered in the FCC-hh, where the formation of the bunch spacing of 5 ns will take place either by debunching in the LHC/PS and rebunching in the LHC/SPS or by building a new injector chain [9, 13].

## A. Introduction to the MSS

The Medium-Stage Synchrotron (MSS) is a high beam-power synchrotron, and the debunching-rebunching method to change the bunch spacing does not work here. A special lattice with a negative momentum compaction factor of $-7.7\times10^{-4}$ has been designed to accelerate proton beams received from its injector, a proton rapid cycling synchrotron (p-RCS), from 10 GeV to 180 GeV with a repetition rate of 0.5 Hz. Then the beams are extracted into the downstream accelerator, a Super Synchrotron (SS). The circumference of the MSS is about 3.5 km, which can accommodate three p-RCS beam batches, each of which contains 112 bunches separated by 25 ns. The SPPC nominal bunch spacing of 25 ns is formed from the bunch structure in the p-RCS with an RF frequency close to 40 MHz. The nominal bunch intensity, defined as the number of particles in one bunch, is $1.5\times10^{11}$ in the MSS when providing beams for the SPPC.

The basic longitudinal dynamics parameters at the MSS injection and extraction are given in Table I, where the bunch-to-bucket transfer match with its injecting synchrotron p-RCS and the receiving synchrotron SS, the space charge effects and so on are taken into account [14]. In addition to the requirements of longitudinal beam dynamics parameters, the following constraints during the process of bunch splitting should be considered: (1) The duration of the whole bunch splitting should be controlled within 0.2 s as short as possible since the repetition period of the MSS is only 2 s. (2) The process of bunch splitting should be implemented under the adiabatic or quasi-adiabatic condition, which requires the voltages of different RF harmonic systems should not change too fast. (3) In the case of being capable of achieving bunch splitting, a relatively lower total RF voltage is preferred due to the expensive cost of RF systems. (4) After bunch splitting, the total longitudinal emittance should not be increased too much, e.g., not exceeding two times as that before splitting.

TABLE I. The basic longitudinal dynamics parameters at the MSS injection and extraction.

| Parameter | Unit | Injection | Extraction |
|---|---|---|---|
| Proton energy | [GeV] | 10 | 180 |
| Ring circumference | [m] | 3478 | 3478 |
| Repetition rate | [Hz] | 0.5 | 0.5 |
| Number of particles per bunch | $10^{11}$ | 1.5 | 1.5 |
| Number of bunches | | 336 | 336 |
| Bunch spacing | [ns] | 25 | 25 |
| Accumulated particles | $10^{14}$ | 0.50 | 0.50 |
| Longitudinal emittance, $4\sigma$ | [eVs] | 0.5 | 1.0 |
| Revolution frequency | [kHz] | 86.3 | 86.3 |
| Circulating beam current | [A] | 0.70 | 0.70 |
| Beam power | [MW] | 0.04 | 0.73 |
| Momentum compaction factor, $\alpha_p$ | $10^{-3}$ | -0.77 | -0.77 |
| RF frequency | [MHz] | 40 | 40 |
| Harmonic number | | 464 | 464 |
| Total RF voltage | [MV] | 3 | 5.8 |
| Synchrotron frequency | [Hz] | 1106.65 | 118.46 |
| Bucket area | [eVs] | 2.37 | 42.77 |
| Bucket half-height, $\Delta E/E$ | $[10^{-3}]$ | 6.80 | 7.43 |
| Full bunch length, $4\sigma_z$ | [ns] | 8.25 | 2.75 |
| Momentum spread, $\sigma_\delta$ | $10^{-3}$ | 1.76 | 0.64 |
| Momentum filling factor | | 0.50 | 0.17 |
| Bunching factor | | 0.22 | 0.07 |

### B. Simulation method for the bunch splitting

#### *1. Introduction to the simulation code -- BLonD*

BLonD (Beam Longitudinal Dynamics) is a simulation code of longitudinal beam dynamics in synchrotrons developed by CERN since 2014 [15]. The code was developed and implemented based on the widely-used Python language, and it has been benchmarked and compared with measurements, theory, or other codes, which enhance its robustness and credibility [16]. It can simulate single-bunch or multi-bunch longitudinal beam dynamics during the acceleration cycle with multiple RF systems, multiple RF stations in different locations, the collective effects, and the implementation of low-level RF controls such as phase noise, phase loop and feedback. It has been applied practically to LHC controlled longitudinal emittance blow-up during the ramp, PS-to-SPS transfer and so on. In this study, it is used to simulate the bunch splitting process.

#### *2. Methodological considerations of different bunch-splitting schemes*

Based on the principal analysis of bunch splitting with multiple harmonic RF systems, as mentioned in Sec. II.B, reasonable RF voltage schemes that will stretch the bunch and split it eventually into multiple bunches need to be carefully studied. For two-fold bunch splitting, one can readily split one bunch into two bunches uniformly with the combination of fundamental harmonic $h_0$ and doubled-harmonic $2h_0$ RF systems, owing to the phase-space symmetry. No complicated RF voltage operation is required except for meeting the adiabatic condition. For three-fold bunch splitting, one can first give a preliminary voltage pattern of the three RF systems with harmonic numbers $h_0$, $2h_0$ and $3h_0$, respectively, then adjust individual RF voltage according to the simulation results of the particle redistribution in the three sub-bunches after splitting. For example, if the middle sub-bunch particles are more than the average, one could lower the $h_0$ voltage, which can help decrease the depth of the potential well in the middle, or increase the $2h_0$ voltage, which can help deepen the potential well on both sides. The method of adjusting the RF voltages is straightforward, and the equally-redistributed sub-bunches can be eventually obtained. However, the RF voltage scheme design for the five-fold bunch splitting becomes much more intricate, which will be explained in detail as follows.

The first fundamental problem is what kinds of harmonic RF systems we indeed require to produce a five-fold bunch splitting. As stated in Sec II.B.2, the RF systems with $h_0$, $2h_0$, $3h_0$, $4h_0$ and $5h_0$ are essential to manipulate five-fold bunch splitting. Here, it is demonstrated why the intermediate RF systems with $2h_0$, $3h_0$ and $4h_0$ are indispensable in addition to the original $h_0$ harmonic and the final $5h_0$ harmonic. In fact, bunch splitting can be regarded as a process of stretching the bunch and then redistributing particles within the lengthened bunch to the newly established high-harmonic RF buckets under the control of multi-harmonic RF voltages. During this process, elongating the bunch by keeping a more-or-less uniform linear density is a key point. The more sub-bunches one bunch will split into, the longer it must be stretched before the splitting. For example, the original bunch needs to be lengthened at least $2\pi/3$ for the triple-fold bunch splitting, whereas it will be at least $6\pi/5$ for the five-fold bunch splitting, as shown in Fig. 4. Otherwise, probably no particles will enter the outermost RF buckets on both sides. To lengthen the bunch that far, an RF harmonic component with its peaks at $\pm 3\pi/5$ is needed. Assume the required RF harmonic component has a harmonic number $kh_0$, where $k$ is an integer, it must satisfy:

$$2 \times \frac{3\pi}{5} \bigg/ \frac{2\pi}{k} \leq \text{even integer, when } k \text{ is odd}, \qquad (6)$$

$$2 \times \frac{3\pi}{5} \bigg/ \frac{2\pi}{k} \leq \text{odd integer, when } k \text{ is even}. \qquad (7)$$

Here, "≤" in Eqs. (6)-(7) means the peak appears at $3\pi/5$ or on its right so that as many particles as the required can enter the outermost stable islands. The results show that $k=3$ can better fulfill this condition. Hence, the RF system with $3h_0$ is vital. Besides, the RF system with $2h_0$ is also crucial to move the core particles outward. Otherwise, the particle number in the final center bunch will be much greater than in the other bunches, and this can be better understood with the potential well. The RF system with $4h_0$ is also required to balance the particle numbers in the outermost bunches and their neighboring bunches. Therefore, all the RF systems with $h_0$, $2h_0$, $3h_0$, $4h_0$ and $5h_0$ are essential to accomplish the manipulation of the five-fold bunch splitting with equal particle redistribution.

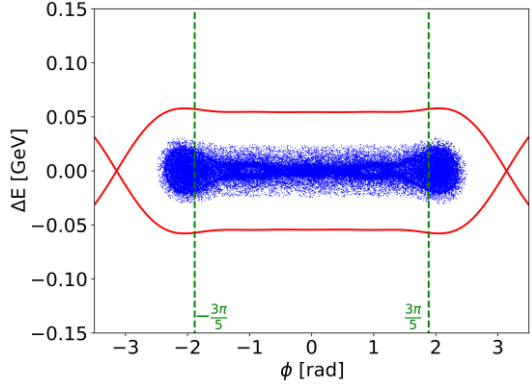

FIG. 4. Requirement for the minimum phase extension for the five-fold bunch splitting. Partial particles within the original bunch must be stretched beyond ±3π/5 so that they can enter the outermost RF buckets on both sides.

Another fundamental problem is the complexity of the particle redistribution with the voltage modulations of multiple harmonic RF systems to achieve the five-fold bunch splitting. For the sake of simplicity, we name the five sub-bunches after the splitting as B1, B2, B3, B4 and B5 from left to right in the phase space plot. Thanks to the symmetrical bunch property, the particle number in B1 and B5 is naturally equal, so as for B2 and B4. Thus, one only needs to consider the equality of B1, B2 and B3. For instance, if the particle numbers are too small in B1 and B2 but too large in B3, the increase of the $2h_0$ voltage can help reduce the particle number in B3 and increase those in B2 and B1, and then the other voltages can be adjusted to balance between B1 and B2. Unfortunately, there is no unique and accurate method to generate the RF voltage programs of different harmonics. One has to optimize the voltage programs of all the five harmonics based on the simulation results.

### *3. Initial RF voltage patterns*

To give the initial RF voltage patterns of the five harmonics for the five-fold bunch splitting, which are the input parameters for simulations with the BLonD code, one can design the voltage curves at some pivotal moments based on the shape of the potential well. Assuming the process of bunch splitting is adiabatic or quasi-adiabatic, the change in the beam emittance shape always keeps pace with the change of the potential well. Hence, the shape of the potential well determines the distribution of the beam in the longitudinal phase space. One can optimize the voltage ratios (a set of *k* values) of different harmonics to obtain the desired potential well based on Eq. (3). Figure 5 shows the potential wells for different sets of *k* values. One can see that the potential well develops from the initial one (in green) that corresponds to the single harmonic first to the flattened one (in red), then further stretched along with the sequential formation of three and four small shallow wells (in blue and orange), finally into five small shallow wells (in purple) with different sets of *k* values in the different moments. During this process, the five harmonic RF systems with $h_0$, $2h_0$, $3h_0$, $4h_0$ and $5h_0$ successively play predominant roles, which makes the potential well wider and wider and redistributes the particles along with the phase continuously. Therefore, these different sets of *k* values can be used as the original RF voltage schemes of the five-fold bunch splitting, with the initial voltage of the fundamental harmonic determined by the momentum filling factor.

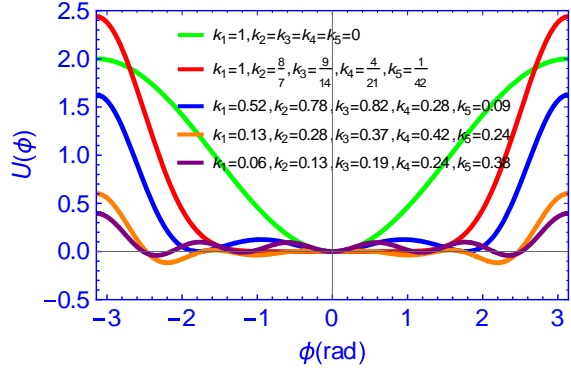

FIG. 5. The shapes of the potential well in the different moments with different combinations of harmonic voltages for the five-fold bunch splitting.

## C. SIMULATION RESULTS

### *1. Five-fold bunch splitting at the extraction energy*

First, we consider the possibility of realizing the five-fold bunch splitting at the extraction energy of 180 GeV. This is natural as one may want to keep the original bunch spacing of 25 ns in the whole acceleration process. The multiple harmonic RF systems with the harmonic number $h_0$ (40 MHz), $2h_0$ (80 MHz), $3h_0$ (120 MHz), $4h_0$ (160 MHz) and $5h_0$ (200MHz) are utilized to achieve the goal. The corresponding RF phases with respect to the bunch center are 0, π, 0, π and 0. In the simulations, 50000 macro-particles representing the real particles of $1.5 \times 10^{11}$ are tracked for about 17400 turns or 0.2 s. The initial bunch follows the two-dimensional Gaussian distribution with the RMS bunch length of 0.688 ns and the RMS absolute energy spread of 0.115 GeV. The intensity-related effects, such as the space charge effects and collective instabilities, were not taken into account in the simulations because they do not significantly affect the implementation of the bunch splitting technique but mainly influence the inner particle redistribution.

With the RF voltage design method in Sec. III.B.3, the initial RF voltage patterns were given at some pivotal moments with the initial voltage of 5.8 MV for the fundamental harmonic RF, as shown in Fig. 6 (a). The final RF voltage of 1.2 MV with $5h_0$ was designed to achieve the perfect match on the longitudinal phase space with the downstream SS. Elaborate optimizations for the RF voltage patterns based on the multi-particle simulations were conducted to obtain more homogeneous bunch populations after the splitting. The optimized RF voltage patterns are presented in Fig. 6 (b). There is no significant difference between the initial RF voltage pattern and the final one after optimization, except that two other moments were added to the optimized one in the fine-tuning effort to obtain better particle redistribution in the final five sub-bunches. This demonstrates that the method of designing the RF voltage patterns based on the potential well is feasible and effective.

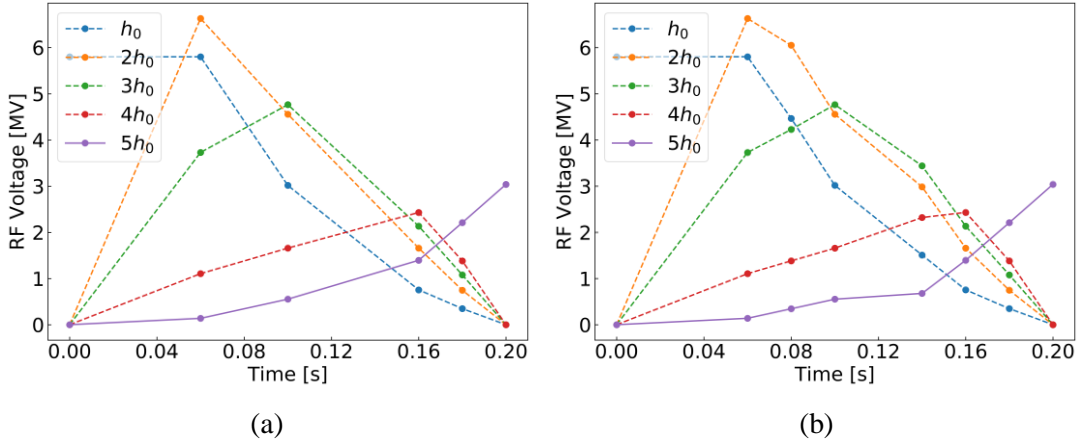

(a)                                                (b)

FIG. 6. The initial RF voltage patterns in (a) and the final one after optimization in (b) for the five-fold bunch splitting at the extraction energy.

Figure 7 represents the evolution process of one bunch splitting into five sub-bunches in the phase space with the optimized RF voltage patterns. One can clearly observe the processes of the bunch stretching and splitting. Actually, the modulations of the different RF harmonic voltages can be divided into two parts: in the first half, the RF systems with $h_0$, $2h_0$ and $3h_0$ predominate, which is similar to triple splitting, and the original bunch evolves into three bunches, see Fig. 7(a)(b)(c); in the second half, with the decrease of the RF voltages with $h_0$, $2h_0$ and $3h_0$ and the increase of the RF voltages with $4h_0$ and $5h_0$, the middle bunch is further lengthened and performs the triple splitting again, and gradually develops into three new sub-bunches, while the other two bunches move outward on the two sides and become the two outermost sub-bunches, see Fig. 7(d)(e). Finally, five equal-spacing sub-bunches are formed under the RF voltage drive with $5h_0$, see Fig. 7(f), which is assisted with $4h_0$ in the process.

The original bunch containing 50000 macro-particles is ultimately divided into five sub-bunches with populations of 9599, 10278, 10170, 10090 and 9863 from B1 to B5 (or from left to right). The maximum population deviation is about 4.0%, which can be acceptable. However, the longitudinal emittance of the five sub-bunches containing 99% macro-particles is 1.333 eVs, 1.215 eVs, 1.169 eVs, 1.225 eVs and 1.342 eVs, respectively, which is much larger than one-fifth of the original longitudinal emittance of 2.424 eVs. This means that the total emittance growth is about 1.59 times. In principle, the longitudinal emittance after the splitting would not blow up significantly under the adiabatic condition. The main cause for the large emittance growth is that the synchrotron tune under the five harmonic RF systems varies dramatically across a much-elongated bunch. Besides, it is difficult to satisfy the adiabatic condition with a given period of 0.2 s and a relatively small synchrotron tune at the extraction energy—the consequent strong filamentation of the emittance results in the large emittance growth.

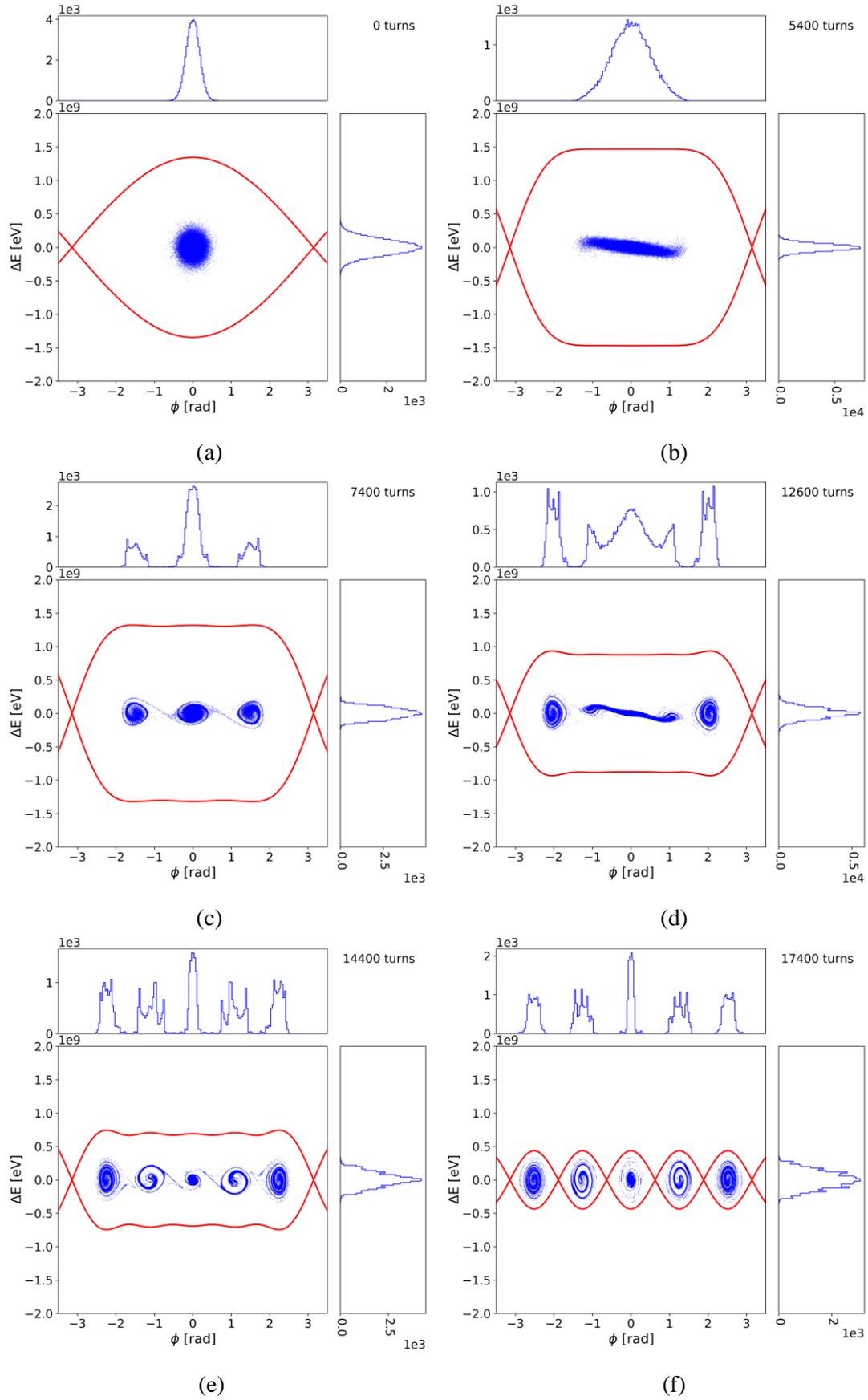

FIG. 7. The bunch distribution in the longitudinal phase space and their projections on the two axes after different turns during the five-fold bunch splitting at the extraction energy. (a) 0th

turn; (b) 5400th turn; (c) 7400th turn; (d) 12600th turn; (e) 14400th turn; (f) 17400th turn.

### 2. Five-fold bunch splitting at the injection energy

As compared with the bunch splitting at the extraction energy, the same manipulation at the injection energy has the advantage of a larger synchrotron tune that is almost 10 times higher than that at the extraction energy and helps meet the adiabatic condition, although the whole acceleration will be executed by the $5h_0$ harmonic RF system instead of the original fundamental harmonic RF system. In this case, the kinetic energy is 10 GeV, the RMS bunch length is 2.06 ns, and the RMS absolute energy spread is 0.019 GeV. Both the initial voltage of the fundamental harmonic RF and the final voltage of the $5h_0$ harmonic RF are set to 3 MV. The RF voltage patterns after optimization and the corresponding splitting results are illustrated in Fig. 8 and Fig. 9, respectively. After splitting, the populations of the five bunches are 10058, 10024, 9828, 10150 and 9940, respectively. The maximum relative deviation in population is around 1.7%. The original longitudinal emittance is 1.194 eVs, and the one after the splitting is 0.246 eVs, 0.267 eVs, 0.261 eVs, 0.268 eVs and 0.252 eVs for the bunches from B1 to B5. The total emittance growth is less than 10%, which is much better than that at the extraction energy.

It is worth pointing out that there are about 872 macro-particles beyond the RF buckets at the end of the splitting, which account for 1.74%, due to the dramatic reduction of the RF bucket area. These particles could be lost during the following acceleration. If the 2D Gaussian distribution for the initial bunch is replaced by a truncated Gaussian of 3 times RMS that almost reflects the usual cases in synchrotrons, so as in the p-RCS, then the number of the potentially lost particles decreases significantly, which only accounts for 0.094%, as shown in Fig. 10. Thus, it reveals that the potentially lost particles mainly come from the beam halo particles of the initial distribution. Besides, the corresponding beam loss rate is about 0.01 W/m, far less than the uncontrolled beam loss standard of 1 W/m for hands-on maintenance [17]. Therefore, the above scheme for the five-fold bunch splitting at the injection energy looks fully acceptable.

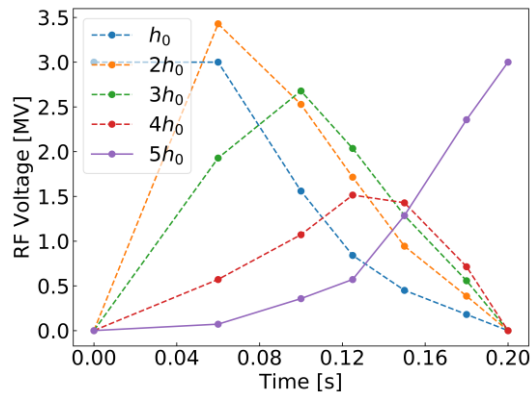

FIG. 8. The RF voltage patterns after optimization for the five-fold bunch splitting at the injection energy.

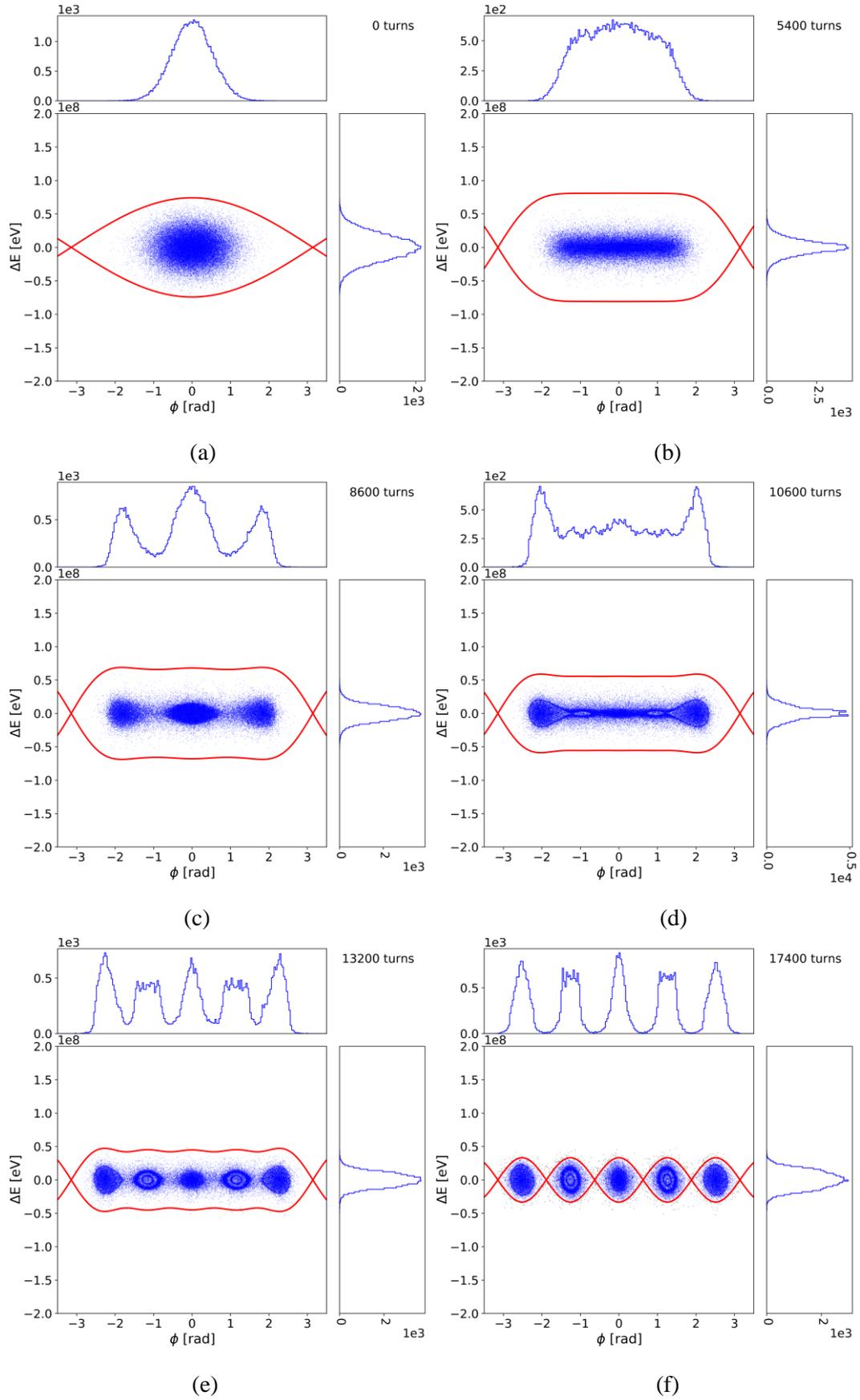

FIG. 9. The particle distribution in the longitudinal phase space and their projections on the two axes after different turns during the five-fold bunch splitting at the injection energy. (a)

0th turn; (b) 5400th turn; (c) 8600th turn; (d) 10600th turn; (e) 13200th turn; (f) 17400th turn. There are about 1.74% of macro-particles beyond the RF buckets at the end of the splitting, which are shown in gray dots in (f).

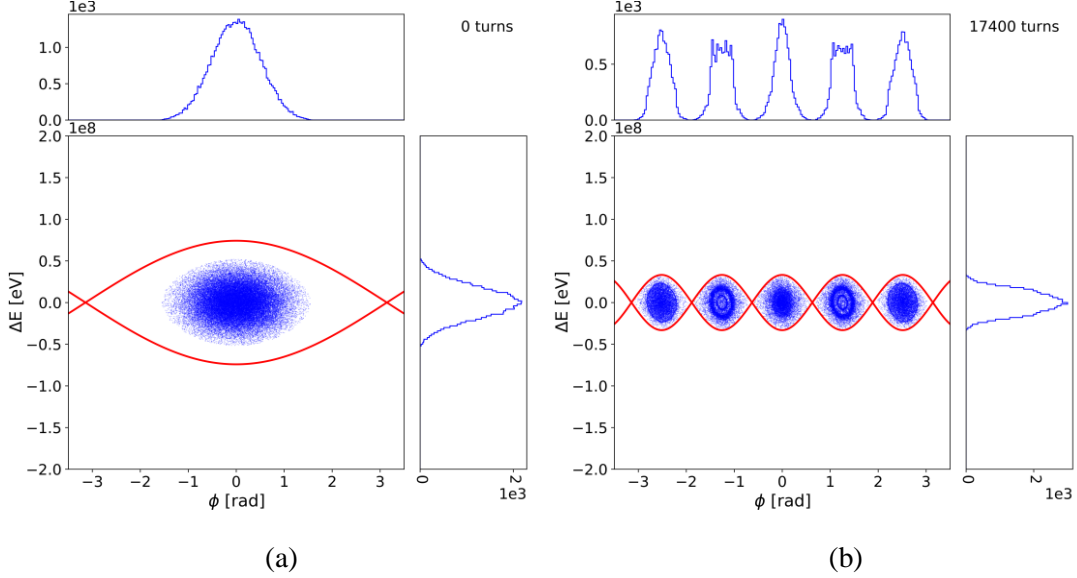

(a)  (b)

FIG. 10. The initial bunch distribution with a Gaussian truncated at 3 times RMS (a) and the final bunch distribution after the five-fold bunch splitting (b). There are only about 0.094% of macro-particles beyond the RF buckets at the end of the splitting.

### *3. Phenomena and result analysis*

The manipulation of the five-fold bunch splitting using the multiple harmonic RF systems can be achieved both at the injection energy and the extraction energy within the limited duration of 0.2 s. However, the phenomena during the bunch splitting obviously become more complicated under five harmonic RF systems than those with the two-fold or three-fold bunch splitting. For instance, the bunch is slightly tilted at the 5400th turn during the bunch lengthening process at the extraction energy but not at the injection energy, see Fig. 7(b) and Fig. 9(b), mainly since the adiabatic condition is better satisfied at the injection energy than at the extraction energy. Besides, the bunch centers of B2 and B4 after the splitting are hollow or sparse for both at the injection energy and at the extraction energy, but more serious for the latter, see Fig. 7(f) and Fig. 9(f). Here, taking the case of the extraction energy as an example, some explanations are given for understanding the phenomena, which is based on the synchrotron frequency distribution in the potential well.

Similar to the beam dynamics in dual harmonic RF systems [18], the Hamiltonian in multiple harmonic RF systems can be given by:

$$H = \frac{Q_{s0}}{2} P^2 + Q_{s0} U(\phi), \tag{8}$$

$$\text{with } P = -\frac{h_0 |\eta|}{Q_{s0}} \frac{\Delta p}{p_0} \text{ and } Q_{s0} = \sqrt{\frac{h_0 e V_0 |\eta|}{2\pi \beta^2 E_0}}. \tag{9}$$

Here, $P$ is the normalized momentum with $\eta$ the slippage factor and $\Delta p/p_0$ the fractional momentum deviation; $Q_{s0}$ is the synchrotron tune at zero amplitude for the fundamental harmonic RF system with $\beta$ the relative velocity and $E_0$ the synchronous energy; $U(\phi)$ is the potential well function, as shown in Eq. (3). For a given trajectory with the maximum phase angle $\hat{\phi}$ in the longitudinal phase space, the corresponding Hamiltonian becomes:

$$H_0 = Q_{s0} U(\hat{\phi}). \tag{10}$$

Moreover, the corresponding action is given by:

$$J(H_0) = \frac{1}{2\pi} \oint P d\phi = \frac{1}{\pi} \int_{-\hat{\phi}}^{\hat{\phi}} P d\phi = \frac{1}{\pi} \int_{-\hat{\phi}}^{\hat{\phi}} \sqrt{2[U(\phi_s) - U(\phi)]} d\phi. \tag{11}$$

Thus, the corresponding synchrotron tune can be calculated numerically by:

$$Q_s = \frac{dH_0}{dJ}. \tag{12}$$

Figure 11 shows the calculated potential well and the corresponding synchrotron tune at the 14400th turn at the extraction energy based on Eqs. (8)-(12). Considering the symmetry, we can merely calculate the synchrotron tune between 0 and π. One can see that the synchrotron tune distribution has several segments, which mainly depends on the number of valleys of the potential well. It is worth noting that the synchrotron tunes at the centers of B2 and B4 at the 14400th turn collapse to zero, whereas the tunes are much larger on both sides of the centers. Therefore, during the formation of B2 and B4, the particles after coming out from the middle bunch will be more prone to oscillate around the centers of B2 and B4 and form a spiral distribution rather than a more compact distribution with the full filamentation effect in a gradually growing bucket. Accordingly, the interiors of the B2 and B4 are somewhat sparse, as shown in Fig. 7(f). The situation gets significantly better at the injection energy, as shown in Fig. 9(f), because the adiabatic condition can be better satisfied. It will certainly result in smaller emittance growth.

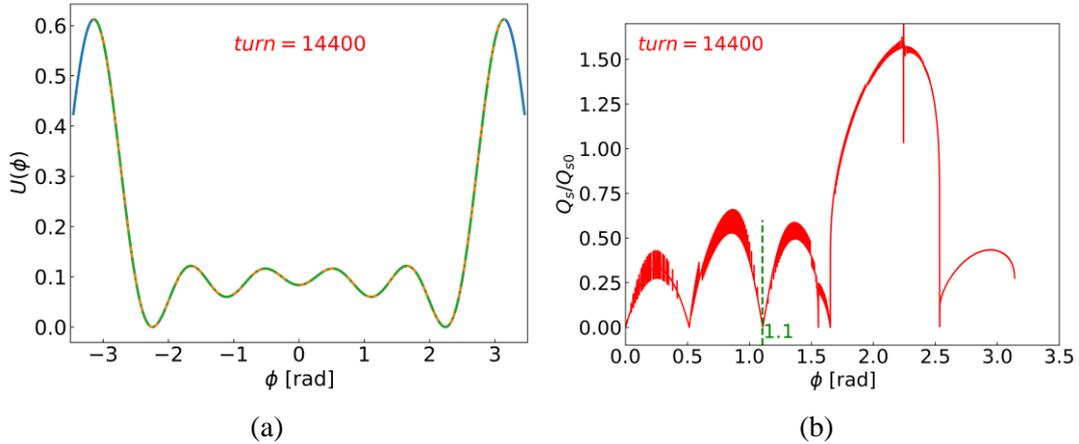

(a)      (b)

FIG. 11. The potential well in (a) and the corresponding synchrotron tune in (b) at the 14400th turn at the extraction energy. One can see that the synchrotron tune at the center of B4 or about 1.1 rad collapses to zero. There are some spikes from numerical calculation errors in the synchrotron tune plot.

Table II summarizes the main properties of the five-fold bunch splitting at the injection energy and the extraction energy. The problem of the longitudinal emittance growth after splitting is considerably mitigated at the injection energy as compared with that at the extraction energy. The main reason is that the synchrotron frequency at the injection is much larger, thanks to the larger slippage factor at the lower energy, so that the adiabatic condition can be better satisfied. Besides, the uniformity of the bunch population after the splitting can be well under control at the injection energy. Although there is some beam loss, it is totally below the tolerable limit, especially when the bunched particles from the p-RCS meet the truncated Gaussian distribution. Therefore, it is better to manipulate the five-fold bunch splitting at the injection energy than at the extraction energy.

TABLE II. The comparison of the simulation results of the five-fold bunch splitting at the injection energy and extraction energy.

| Schemes | Beam loss rate | Max. deviation in the population | Total longitudinal emittance growth |
| --- | --- | --- | --- |
| Injection - Gaussian | 1.74% | 1.7% | 8.4% |
| Inj. - truncated Gaussian | 0.094% | 1.4% | 9.1% |
| Extraction - Gaussian | 0 | 4.0% | 159% |

### IV. Conclusions and discussion

In this paper, the general multi-fold bunch splitting technique with multi-harmonic RF systems is summarized, and it is the first time to study the manipulation of five-fold bunch splitting. This method involves the selection of the number of harmonic RF systems, the determination of the relative phases of different harmonics, and the design and optimization of each harmonic voltage that are the most important and challenging. The method of designing the RF voltage patterns at some pivotal points based on the potential well under the adiabatic condition is first applied, and the further optimization of the voltages is based on the multi-particle simulations. Finally, the synchrotron tune distribution across the bucket in the multi-harmonic RF systems has been numerically calculated to understand the complicated phenomena related to the particle redistribution during the splitting.

The five-fold bunch splitting method using multi-harmonic RF systems has been successfully applied to the MSS, the third stage of the SPPC injector chain, to obtain a bunch spacing of 5 ns from the original 25 ns. The results show that the bunch population redistribution after the splitting can be well controlled both at the injection energy and extraction energy, but the control over the longitudinal emittance growth is significantly better with the former. The main reason for the difference is that the adiabatic condition is much better satisfied at the injection energy within a limited period of 0.2 s. With the bunch population deviations of less than 1.7% and the total longitudinal emittance growth of less than 10%, the scheme for the five-fold bunch splitting at the injection energy is considered promising and acceptable. Certainly, the method can also be applied to other high-intensity proton synchrotrons or other accelerators requiring multi-fold bunch splitting.

In the future, further studies are needed to make the five-fold bunch splitting method fully applicable in the SPPC. The effects related to beam intensity, like the space charge effects, the beam loading effects and the collective instabilities, should be considered, as they will aggravate the emittance growth and change the uniformity of bunch redistribution and even cause beam loss.

## ACKNOWLEDGMENTS

The authors would like to thank Alexandre Lasheen of CERN, one of the BLonD developers, for beneficial help and guidance for the use of the BLonD code. The study is supported by the National Natural Science Foundation of China (11575214, 12035017) and CAS-IHEP Fund for PRC\US Collaboration in HEP.

## REFERENCE

[1] Michiko G. Minty, and Frank Zimmermann, Beam Techniques--Beam Control and Manipulation, in *Proceedings of Lectures given at the US Particle Accelerator School*, *June 14-25, 1999*; Report No. SLAC-R-621, April 2003. https://doi.org/10.2172/813024.

[2] S.Y. Lee, Accelerator Physics (World Scientific, Singapore, 2004), pp. 317-340.

[3] R. Garoby, RF gymnastics in synchrotrons, in *Proceedings of CERN Accelerator School 2010—RF for accelerators*, *Ebeltoft, Denmark, 2010*, arXiv:1112.3232.

[4] R. Garoby, Bunch merging and splitting techniques in the injectors for high energy hadron colliders, in *17th International Conference on High Energy Accelerators (HEACC'98), Dubna, Russia, 1998,* pp.172-174; Report No. CERN-PS-98-048-RF, http://cds.cern.ch/record/367499.

[5] R. Garoby, S. Hancock, and J.L. Vallet, Demonstration of bunch triple splitting in the CERN PS, in *7th European Accelerator Conference, Vienna, Austria, 2000*; Report No. CERN-PS-2000-038-RF, http://cds.cern.ch/record/453506.

[6] R. Garoby, Multiple bunch-splitting in the PS: results and plans, Report No. CERN-PS-2001-004-RF, Geneva, Switzerland, 2001, http://cds.cern.ch/record/488252.

[7] R. Gamage and T. Satogata, Bunch splitting simulations for the JLEIC ion collider ring, in *Proceedings of the 7th International Particle Accelerator Conference, Busan, Korea, 2016*, pp. 2448-2450, https://doi.org/10.18429/JACoW-IPAC2016-WEPMW013.

[8] The CEPC-SPPC study group, CEPC Conceptual Design Report, Volume I-Accelerator, Report No. IHEP-CEPC-DR-2018-01, IHEP-AC-2018-01, 2018, Chap. 8, arXiv:1809.00285.

[9] Abada, A., Abbrescia, M., AbdusSalam, S.S. et al., FCC-hh: The Hadron Collider. Eur. Phys. J. Spec. Top. 228, 755–1107 (2019). https://doi.org/10.1140/epjst/e2019-900087-0.

[10] R. Garoby, PS Machine Development Report Debunching (h=20) and Rebunching (h=84) at 26 GeV in the PS, Report No. PS/RF/Note 98-02 (MD), 1998. http://cds.cern.ch/record/957830/files/cer-002624160.pdf.

[11] A. Hofmann and S. Myers, Beam dynamics in a double RF system, in *Proceedings of 11th International Conference on High-Energy Accelerators, CERN, Geneva, Switzerland, 1980*, pp.610-14; Report No. CERN-ISR-TH-RF /80-26, http://cds.cern.ch/record/879237.

[12] L.J. Wang and J.Y. Tang, "Luminosity optimization and leveling in the Super Proton–Proton Collider", Radiat Detect Technol Methods (2021). https://doi.org/10.1007/s41605-020-00233-6.

[13] H. Bartosik, H. Damerau, and E. Shaposhnikova, Implications of 5 ns bunch spacing for the FCC-hh injector chain, in *Future Circular Collider Week 2017, Berlin, Germany, 2017*, https://indico.cern.ch/event/556692/contributions/2567943/attachments/1468210/2271031/spacing5nsFCC_final.pdf.

[14] L.H. Zhang, J.Y. Tang, Y. Hong, Y.K. Chen and L.J. Wang, Optimization of Design Parameters for the SPPC Longitudinal Dynamics, arXiv:2101.10623.

[15] CERN Beam Longitudinal Dynamics code BLonD, http://blond.web.cern.ch.

[16] H. Timko, J. Esteban Müller, A. Lasheen, and D. Quartullo, Benchmarking the beam longitudinal dynamics code BLonD, in *Proceedings of the 7th International Particle Accelerator Conference, Busan, Korea, 2016,* pp.3094-3097, https://doi.org/10.18429/JACoW-IPAC2016-WEPOY045.

[17] J.Y. Tang, Rapid cycling synchrotrons and accumulator rings for high-intensity hadron beams, Reviews of Accelerator Science and Technology, Vol. 6, (2013), https://doi.org/10.1142/S1793626813300077.

[18] J.Y. Liu, D.D. Caussan, M. Ellison, S.Y. LEE, D. LI, A. Riabko and L. Wang, Analytic solution of particle motion in a double rf system, Part. Accel. **49** (1995) pp.221-251, http://cds.cern.ch/record/1120213.